\documentclass[a4,11pt]{article}
\usepackage{epsfig}
\usepackage{amssymb}
\usepackage{amsmath}
\usepackage{amscd}
\usepackage[english]{babel}
\begin{document}
\newcommand{\cts}{\cos{\theta^*}}
\newcommand{\ct}{\cos{\theta}}
\newcommand{\sts}{\sin{\theta^*}}
\newcommand{\st}{\sin{\theta}}
\newcommand{\demi}{\frac{1}{2}}
\newcommand{\GeV}{\text{GeV}}
\newcommand{\eV}{\text{eV}}
\newcommand{\MeV}{\text{MeV}}
\newcommand{\cs}[1]{\langle\sigma_{#1}\rangle}
\newcommand{\cm}{\text{cm}}

\title{Kinematics of an off-axis neutrino beam}
\author{Jean-Michel Levy \thanks{Laboratoire de Physique Nucl\'eaire et de Hautes Energies,
CNRS - IN2P3 - Universit\'es Paris VI et Paris VII, Paris.   \it Email: jmlevy@in2p3.fr}}
\maketitle
\pagenumbering{arabic}
\sloppy
\maketitle
\begin{abstract}
We review the kinematics of a neutrino beam in the idealized case where 
the parent mesons momenta are parallel, but without any other 
approximation. This reveals several interesting features, in particular in the
off-axis case,  which are hidden by the approximations made in a previous treatment. 
\end{abstract}
\newpage
\section{Introduction.}
The kinematics of an off-axis neutrino beam in the approximation where the parent meson beam is 
divergenceless was posed as a problem to the student in these Archives a long time ago.
However, the author of \cite{McD} seems not to have noticed that the problem is amenable to an exact 
treatment which reveals interesting aspects hidden by the approximations he uses.
In this note, we show how to implement exactly the Lorentz transformation in the differential
(angle, energy) distribution and we exhibit some features of the results.\\

Among others, it is shown that a sizeable part of the flux at a given lab angle comes from neutrinos going backwards in the center of mass frame of 
the decaying meson. Given the approximations made in \cite{McD}, it is not clear that they are taken into account in the results presented. 
How this might have a bearing on the calculations of neutrino fluxes being performed with more realistic divergent meson beams in the preparation of 
upcoming experiments is unknown to the present author.

\section{Conventions}
Starred variables refer to the c.o.m. frame (com) of the decaying meson.\\
The neutrino energy in the lab is $E$ and its angle with respect to the beam axis is $\theta$ in the lab and $\theta^*$ in the decaying meson frame.
For illustrative purposes, we use a $30~ \GeV$ proton beam energy corresponding to the T2K experiment  and a 
pion spectrum of the form $(E_{p}-E_{\pi})^5$  as done in \cite{McD}

\section{Neutrino generating decays}

$\pi^{\pm} \rightarrow l^{\pm} + \nu_l \;\; (\bar{\nu}_l)$\\
or\\
$K \rightarrow \pi + l \;\; (\bar{l}) + \nu_l \;\; (\bar{\nu}_l)$\\

with $l = \mu \;\;\text{or}\;\; e$ and \\

$\mu \rightarrow \nu_{\mu} + e + \bar{\nu_e} $\\

In the second and third cases, $E^*$ which is henceforth the neutrino com energy is
not fixed by the conservation laws, but it makes little difference in what follows,
except that the lab energy is not rigidly related to the lab angle and meson velocity.\\
Also note that for $\mu$ decays, the divergenceless hypothesis does not hold since $\mu$'s
are themselves produced in meson decays. The forthcoming analysis should therefore be 
amended in this case.\\

For the two-body case, we take $\pi \rightarrow \mu \; \nu$ as our paradygm:\\
$E_{\mu} + E_{\nu} = m_{\pi}$ and $p_{\mu} + p_{\nu} = 0$ together with $m_{\nu}=0$ imply that\\
$E_{\nu} = E^* = \frac{m_{\pi}^2-m_{\mu}^2}{2m_{\pi}}$ and $p^* = E^* = 29.8~\text{MeV}$\\

\section{Lorentz transformation}
We take the decaying meson direction as polar axis in the following, and consider the meson beam as divergenceless.\\
Consequently we shall not make use of the azimuthal degree of freedom and reduce the problem to two dimensions.\\

The neutrino energy-momentum vector in com is therefore $E^*(1,\cts,\sts)$\\

Let $\gamma = \frac{E_{\pi}}{m_{\pi}}$ be the $\pi$ Lorentz factor and $\beta = \frac{p_{\pi}}{E_{\pi}}$
its velocity for $c=1$\\

The lab neutrino energy-momentum vector is given by:\\

\begin{eqnarray}
E\qquad\;\, &=& \gamma E^*(1+\beta\cts)  \label{en} \\
E~\ct &=& \gamma E^*(\cts + \beta)\\ 
E~\st &=& E^*\sts   
\end{eqnarray}

We set $p=E$ since our neutrinos are evidently considered as massless here. This is not an approximation
when only these kinematics are considered. To see, it is enough to remember that the neutrino Lorentz
$\gamma$ in the $\pi$ rest frame is $3~10^7$ for a $m = 1~\eV/c^2\;$ neutrino, whilst the $\gamma$ of a $139~\GeV$
$\pi$ is $10^3$

From this follows:

\begin{eqnarray}
\ct & = & \frac{\cts + \beta}{1+\beta\cts}   \label{ct} \\
\st & = & \frac{\sts}{\gamma(1+\beta\cts)}   \label{st}
\end{eqnarray}

All these equations are inversed by the substitution $\beta \leftrightarrow -\beta$
and $\theta \leftrightarrow \theta^*$\\

The relation for $\tan{\theta}$ following from (\ref{ct} and \ref{st}) has no interest. 
However, a complete picture of the relation $\theta \leftrightarrow \theta^*$ is obtained 
by using $\tan{x/2} = \frac{\sin{x}}{1+\cos{x}}$ which yields:

\begin{eqnarray*}
\tan{\theta/2} = \sqrt{\frac{1-\beta}{1+\beta}}\tan{\theta^*/2}
\end{eqnarray*}
which shows that one angle is a monotonous function of the other on the complete
$[0,\pi]$ interval, notwithstanding the well known headlight effect.  \\

The following useful relations should also be noted:

\begin{eqnarray}
(1-\beta\ct)(1+\beta\cts)\gamma^2 &=& 1 \label{mag}\\
\text{which entails:~}\qquad E &= &\frac{E^*}{\gamma(1-\beta\ct)} \label{en2}\\
\text{and the inverse of (\ref{ct}):}\qquad \cts &=&\frac{\ct - \beta}{1-\beta\ct} \label{inv}
\end{eqnarray}
\section{Kinematical limits}

From equation (\ref{st})  one sees that $\st = \frac{~E^*}{E}\sts$ and therefore 
\begin{eqnarray*}
\st &\leq& \frac{~E^*}{E}
\end{eqnarray*}
which shows that a given neutrino energy $E$ can only be found up to a maximum lab angle
\begin{eqnarray}
\theta_{\max}(E) = &\arcsin{\frac{~E^*}{E}} \approx &\frac{30~\MeV}{E}  
\label{thm} 
\end{eqnarray}
which is small for most neutrinos.\\

Conversely, at a given angle $\theta$ from the (supposedly divergenceless) beam, the maximum 
neutrino energy is:
\begin{eqnarray}
E_{\max}(\theta)& = &\frac{E^*}{\st} \label{emx}
\end{eqnarray}

This bound is valid, of course, for not too small angles. Using (\ref{en2}), we can retrieve it more accurately by calculating
$\frac{\partial E}{\partial \gamma}$ at fixed $\ct$ to get:
\begin{eqnarray*}
\frac{\partial E}{\partial \gamma} = \frac{E}{\beta\gamma}\frac{\ct - \beta}{1-\beta\ct} 
\end{eqnarray*}

\begin{figure}[h]
\begin{center}
\epsfig{file=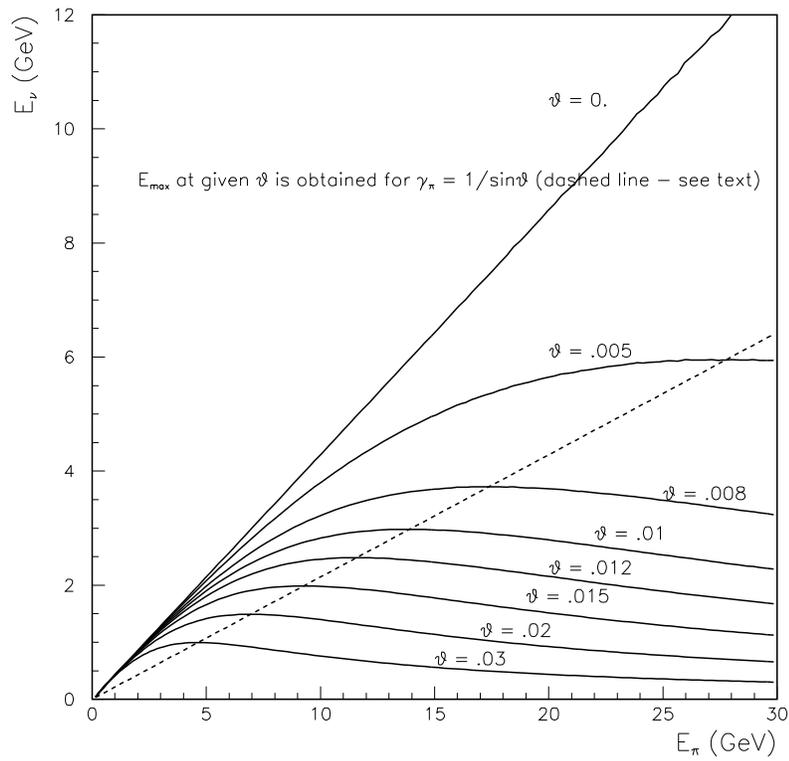,width = .9\linewidth} 
\caption{Illustration of the $E_{\max}(\theta)$ bound} \label{fig: enufpi}
\end{center}
\end{figure}

It follows that the maximum lab energy at fixed $\ct$ is obtained for $\beta = \ct$. This correspond to $\gamma = \frac{1}{\st}$ and
we find, using again (\ref{en2}), that $E_{\max} = E^*\gamma = \frac{E^*}{\st}$ in conformity with (\ref{emx}). These relations are illustrated on 
Fig.~\ref{fig: enufpi}

So, although the bound seems to be  independant of the beam, it makes sense only if there exist pions with 
$E_{\pi} > m_{\pi}/\st$ and (\ref{emx}) must be complemented by the restriction $\st > \frac{1}{\max{\gamma_{\pi}}}$ \\

The stationnarity of $E_{\nu}(\gamma_{\pi})$ will result in many neutrinos emitted by pions around that value of $\gamma$ to pile up 
at the same lab energy $E_{\max}$. Since they come from pions with $\gamma$'s both below and above $1/\st$,
they are emitted both forwards and backwards in the pion frame. We will find the distributions of both components presently.\\
 
For larger pion energies, neutrinos at angle $\theta$ with respect to the axis are emitted more and more backwards in com 
with the result that their lab energy slightly decreases with $E_{\pi}$ (see Fig.~\ref{fig: enufpi})\\

 
In an analogous manner, we can retrieve the bound (\ref{thm}) by using (\ref{en2}) to calculate the derivative 
$\frac{\partial \ct}{\partial \gamma}$ at fixed $E$. We find: 
\begin{eqnarray*}
\frac{\partial \ct}{\partial \gamma} = \frac{1}{\beta\gamma}(1-\frac{\ct}{\beta})
\end{eqnarray*}
and the minimum $\ct$ at fixed lab energy is $\ct = \beta$ so that again, the stationnarity wrt $\gamma$ will result in a piling up around $\ct$ of 
neutrinos coming from pions with energies around $m_{\pi}/\st$. This is illustrated by the distributions (derived below) plotted on Fig.~\ref{fig: cdist}

\section{Probability distributions}

\subsection{Fixed meson energy}
In this subsection, we consider that the decaying meson velocity is fixed.
Therefore there is but one final state variable for a 2-body decay (not counting the azimuth)
and the lab angular distribution is completely determined even in the case of 3-body decay
of the parent meson. 
\subsubsection{Angular distribution}
By differentiating (\ref{inv}) with respect to $\ct$, we find 
\begin{eqnarray} 
\frac{\partial \cts}{\partial \ct} &= &\frac{1-\beta^2}{(1-\beta\ct)^2}
\end{eqnarray}

Since the meson is spinless it decays isotropically in com
and the $\cts$ distribution is flat on the interval $[-1,1]$

The lab distribution of the cosine for a given meson energy is thus:

\begin{eqnarray}
\frac{\partial P}{\partial \ct} &= &\demi \frac{1-\beta^2}{(1-\beta\ct)^2}\\
&&\makebox{\rm{or, re-establishing the azimuthal degree of freedom}} \nonumber \\
\frac{\partial P}{\partial \Omega} &= & \frac{1}{4\pi\gamma^2(1-\beta\ct)^2}  \label{dbl}
\end{eqnarray} 

This distribution is valid without change for 3-body decays.
\subsubsection{Energy-angle distribution}
Equations (\ref{en}) and (\ref{ct}) allow to calculate all the necessary
partial derivatives:\\
\begin{eqnarray*}
\frac{\partial E}{\partial E^*} \;\; = &\gamma(1+\beta \cts) \qquad &\frac{\partial \ct}{\partial E^*} \;=\; 0\\
\frac{\partial E}{\partial \cts} = &\gamma\beta E^* \qquad \qquad \qquad & \frac{\partial \ct}{\partial \cts} = \frac{1}{\gamma^2(1+\beta\cts)^2}
\end{eqnarray*}

from which we get the Jacobian:
\begin{eqnarray*}
\frac{D(E^*,\cts)}{D(E~,\ct~)} &= \gamma(1+\beta\cts) &= \frac{1}{\gamma(1-\beta\ct)}
\end{eqnarray*}

So assuming some energy distribution $F(E^*)$ for $E^*$ (independant of the angles by isotropy)
we get:
\begin{eqnarray}
\frac{\partial P}{\partial E \partial \Omega} = \frac{\partial P}{\partial E^* \partial \Omega^*} 
\frac{1}{\gamma(1-\beta\ct)} = F(\gamma E(1-\beta\ct))\frac{1}{4\pi\gamma(1-\beta\ct)}
\end{eqnarray} 

For 2-body meson decays, there is only one variable ($F$ is a Dirac $\delta$) and this reduces to the angular 
distribution given in (\ref{dbl}) after elimination of $\delta$; the result can alternatively be expressed as a 
distribution in energy.\\

For 3-body meson decays, $F$ must be found from the decay dynamics. From (\ref{en}) and (\ref{mag}) $E_{\max}$ for a 
given angle is given by $\frac{E^*_{\max}}{\gamma(1-\beta\ct)}$ with \\$E^*_{\max} = 
\frac{M^2_K-(m_{\pi}+m_l)^2}{2M_K}$ obtained when the mass of the system recoiling against the neutrino in com is 
minimal.
\subsection{Introducing a meson spectrum}
We no longer consider that the energy of the parent meson is fixed. We assume an energy distribution, say $g(E_{\pi})$
for the $\pi$ mesons in the lab. Note that we use '$\pi$' for definiteness. When it comes to 3-body decays, $K$
would be more appropriate.\\

Even assuming constant $E^*$ (2-body decay), the two final state variables $E$ and $\ct$ can now vary separately. To 
derive their distribution, we need the Jacobian of the transformation $(E_{\pi},\cts) \rightarrow (E,\ct)$\\

\subsubsection{Transforming $E_{\pi},\cts \rightarrow E,\ct$}

When performing this transformation, $E^*$ must be considered as a fixed parameter. In the case of 3-body decays,
the $(E, \ct)$ distribution found below will be simply multiplied by the $E^*$ com distribution and integrated over 
this variable.\\

It is much simpler to calculate the partial derivatives of the final variables with respect to $E_{\pi}$ and 
$\cts$ than the contrary. All we need is again contained in the two relations (\ref{en}) and (\ref{ct}) and the 
relation 
\begin{eqnarray}
\frac{d \beta}{d \gamma} = \frac{1}{\beta\gamma^3}
\end{eqnarray}
The necessary derivatives are as follows:

\begin{eqnarray*}
\frac{\partial E}{\partial E_{\pi}} = \frac{E^*}{m_{\pi}}(1+1/\beta\cts)\qquad && \frac{\partial E}{\partial \cts} = 
E^*\gamma\beta\\
\frac{\partial \ct}{\partial E_{\pi}} = \frac{\sts^2}{(1+\beta\cts)^2}\frac{1}{m_{\pi}\beta \gamma^3}&& \frac{\partial \ct}{\partial \cts} = 
\frac{1}{\gamma^2(1+\beta\cts)^2}
\end{eqnarray*}

Therefore we get:

\begin{eqnarray*}
\frac{D(E~,\ct)}{D(E_{\pi},\cts)} = \frac{E^*}{m_{\pi}\gamma^2\beta}\frac{\cts}{1+\beta\cts}
\end{eqnarray*}
Using the relations established above, the inverse Jacobian reduces to:

\begin{eqnarray*}
\frac{D(E_{\pi},\cts)}{D(E~,\ct)} = \frac{m_{\pi}\beta}{E^*}\frac{1}{(\ct-\beta)}
\end{eqnarray*}
which, not surprisingly, is singular when $\ct = \beta$, corresponding to $\cts = 0$\\

\subsubsection{Inverse transformation}

To get the final $N(E,\ct)$ distribution for given $E^*$, we must invert the relation yielding $(E,\ct)$ as a function
of $(E_{\pi}, \cts)$. This is done as follows:\\

From (\ref{ct}) and (\ref{st}) one easily deduce the following equations:

\begin{eqnarray}
\gamma &=& \frac{1-\ct\cts}{\st\sts}  \label{gm1}\\
\beta &=& \frac{\ct-\cts}{1-\ct\cts} \label{be1}
\end{eqnarray}

We further define: $r \equiv \frac{E}{~E^*}$ so that (\ref{st}) reads $\sts = r\st$\\

Hence (\ref{gm1}) and (\ref{be1}) become: 
\begin{eqnarray}
\gamma &=& \frac{1-\ct\epsilon\sqrt{1-r^2\st^2}}{r\st^2} \label{gm2}\\
\beta &=& \frac{\ct-\epsilon\sqrt{1-r^2\st^2}}{1-\ct\epsilon\sqrt{1-r^2\st^2}} \label{be2}
\end{eqnarray}
with $\epsilon = \pm 1$ according to whether $\cts > 0$ or $\cts <0$, that is, $\ct > \beta$ or $\ct < \beta$ (cf. (\ref{inv}));
observe that (\ref{be1}) implies that $\cts$ is always smaller than $\ct$. Other constraints on $\beta$ and $\gamma$
are automatically obeyed by the expressions (\ref{gm1}) and (\ref{be1})\\

We also get from (\ref{be2}) 
\begin{eqnarray}
\frac{\ct -\beta}{\beta}&=& \frac{\st^2\epsilon\sqrt{1-r^2\st^2}}{\ct-\epsilon\sqrt{1-r^2\st^2}}  \label{cmb}
\end{eqnarray}

This allows us to rewrite the Jacobian in terms of laboratory neutrino variables as:
\begin{eqnarray}
\frac{D(E_{\pi},\cts)}{D(E~,\ct)} = \frac{m_{\pi}}{E^*}\frac{\ct-\epsilon\sqrt{1-r^2\st^2}}{\st^2\sqrt{1-r^2\st^2}}
\end{eqnarray}


To see when each of the two {\it a priori} possible values of $\cts$ apply, we observe first that (\ref{en}) and (\ref{en2}) 
show that the physical range of $r$ is 
$$[ \gamma(1-\beta)=\sqrt{\frac{1-\beta}{1+\beta}} \rightarrow \gamma(1+\beta)=\sqrt{\frac{1+\beta}{1-\beta}}]$$ 

Further, these two equations allow us to write $\ct$ and $\cts$ as functions of $r$ and the boost parameters as follows:
\begin{eqnarray*}
&&\cts  = \frac{r-\gamma}{\beta\gamma} \label{ctrg}\\
&&\ct\,\, = \frac{r\gamma-1}{\beta\gamma r} \label{ctsrg}
\end{eqnarray*} 


The expression for $\cts$ solves the $\epsilon$ problem: 
\begin{itemize}
\item for $r > 1$ the two signs of $\cts$ are possible since we can (if the beam permits !) have $\gamma <r$ and $\gamma >r$
The first case corresponds to a lower energy meson with the neutrino going forward in its rest frame, whilst the second means
a more energetic meson needed to compensate the backward projection of the neutrino in this frame.
\item for $r = 1$ there is but one possibility left, corresponding to \\ $\ct = -\cts = \frac{\gamma-1}{\beta\gamma}$
\item for $r < 1$, only the $\cts < 0$ solution exists. When $r = 1/\gamma$, $\ct = 0$ 
\item $\sqrt{\frac{1-\beta}{1+\beta}} \le r \le \sqrt{1-\beta^2}$ corresponds to 
neutrinos going backwards in the lab.
\end{itemize}

This is where we diverge from the treatment of \cite{McD}. Equation $(18)$ of this author seems to imply that only 
the $\cts>0$ contribution is taken into account. However, $\cts=0$ corresponds to $\theta = \arccos{\beta}$ which is 
generally very small, so that a sizeable number of almost forward neutrinos can be forgotten here. Fig.~\ref{fig: endist1}
strikingly illustrates this remark.\\

Therefore, to get the joint distribution for $(\ct, E)$ we must add the contributions of these two solutions 
when both apply. The joint distribution for $E_{\pi}$ and $\Omega^*$  which is simply $\frac{g(E_{\pi})}{4\pi}$ 
is transformed to:


\begin{multline}
\frac{\partial^2 P}{\partial E \partial \Omega} = \frac{m_{\pi}}{4\pi~E^*\st^2\sqrt{1-r^2\st^2}}\times \label{fin} \\[2mm]
\left [ g(m_{\pi}\gamma_+)|\ct-\sqrt{1-r^2\st^2}|+g(m_{\pi}\gamma_-)|\ct+\sqrt{1-r^2\st^2}|\right ] 
\end{multline}
where $\gamma_{\pm}$ means (\ref{gm2}) with $\epsilon = \pm 1$ and {\it only the second term is to be kept if $r \le 1$} .\\

This must be multiplied by the probability density of $E^*$ and integrated over it to get the final distribution in ($\ct,E$). In the case of a 2-body 
decay, this results in the mere replacement $E^* \rightarrow E^*_0$ in (\ref{fin}), where the value of $E^*_0$ stems from the conservation laws.

\begin{figure}[ht]
\begin{center}
\epsfig{file=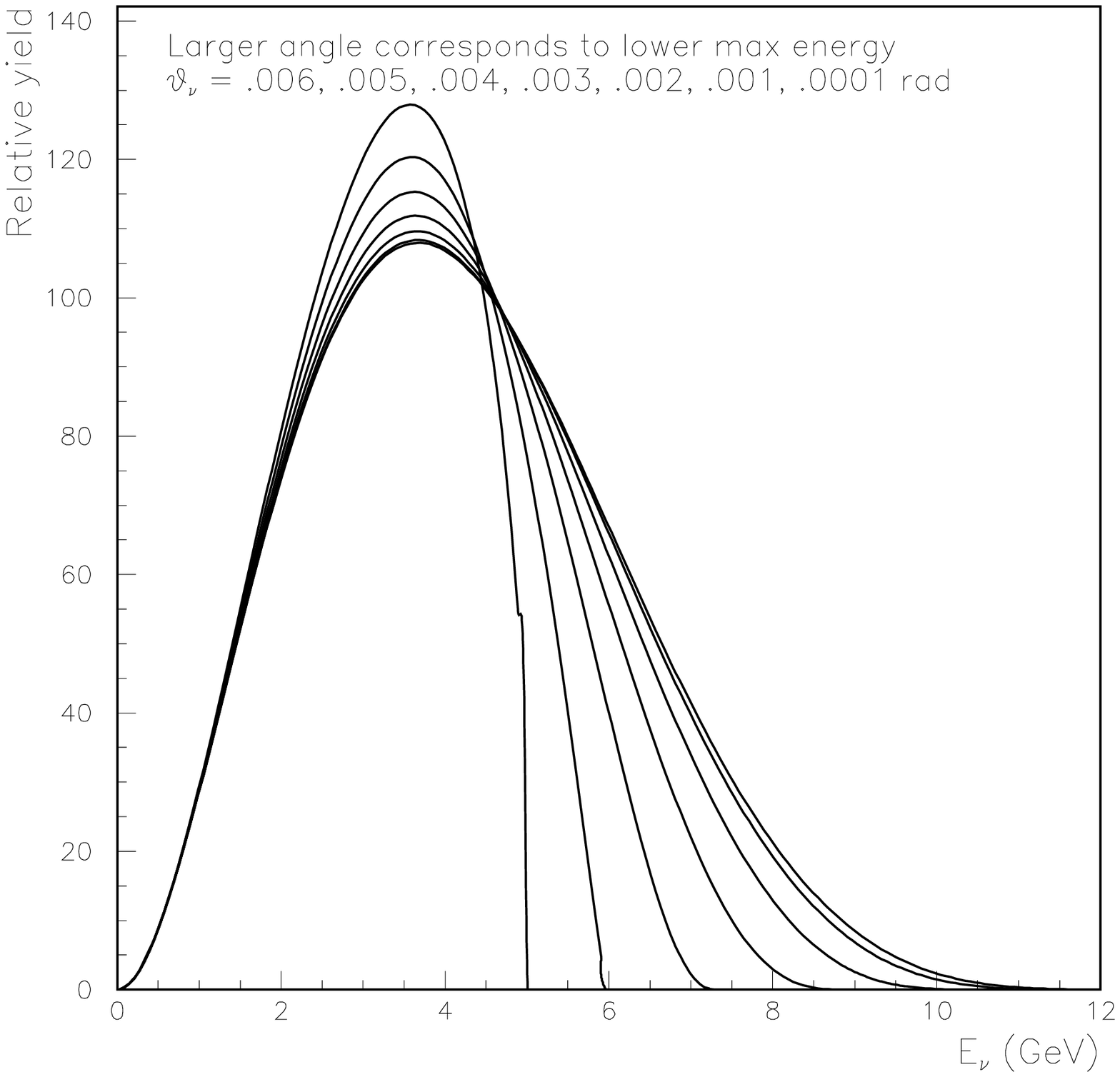,width= .9\linewidth}
\caption{Distribution of $E_{\nu}$ at fixed angle} \label{fig: endist2}
\end{center}
\end{figure}

\begin{figure}[ht]
\begin{center}
\epsfig{file=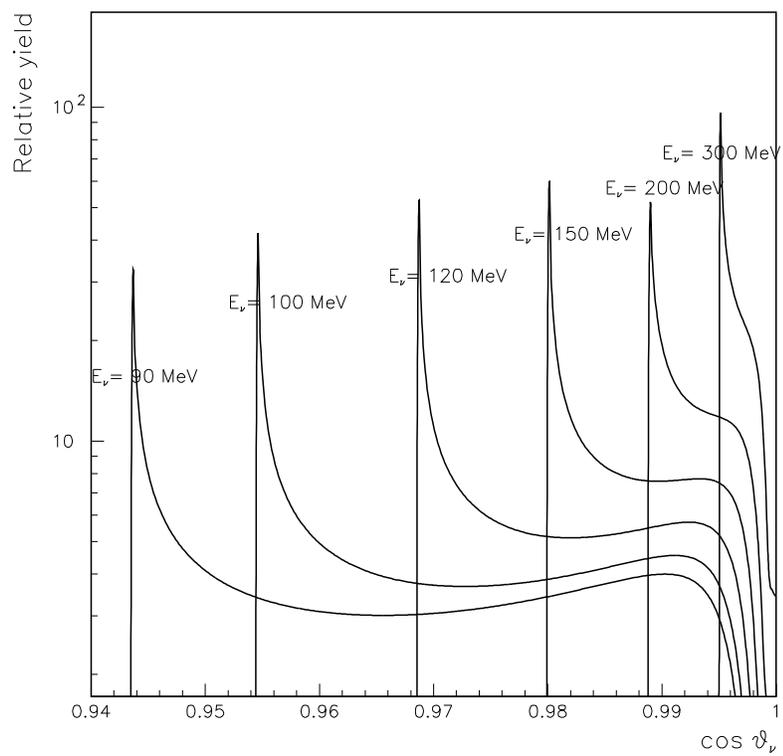,width= .9\linewidth}
\caption{Distribution of $\ct_{\nu}$ at fixed neutrino energy} \label{fig: cdist}
\end{center}
\end{figure}

Figures~ \ref{fig: endist1}, \ref{fig: endist2} and \ref{fig: cdist} show the behaviour expected for constant $E$ or constant $\ct$. The last 
displays Jacobian peaks at $\theta =\arcsin{E^*/E}$ corresponding to the limits explicited in (\ref{thm}). The first shows the same phenomenon for 
fixed $\theta$ with piling up at $E = E^*/\sin{\theta}$ in conformity with (\ref{emx}) The effect seems to disappear (Fig~\ref{fig: endist2}) for 
$\theta = .006$. As explained after equation (\ref{emx}), the limit (in $E$) and the concomitant Jacobian peak exists only if pions of energy larger 
than $m_{\pi}/\st$ are present in the beam. With $E_p = 30~\GeV$ and the spectrum shape adopted here, this corresponds to $\theta > 4.6~10^{-3} $. 
However, the pion spectrum decreases sharply before this limit. \\Note that although the contribution of neutrinos going backwards in the pion rest 
frame dies off when going to smaller angles due to the scarcity of energetic pions, they concentrate at $E^*/\st$, reinforcing the peak which exists 
already in the distribution of neutrinos going forward in this frame.


\section{Conclusions} 
\begin{figure}[ht]
\begin{center}
\epsfig{file=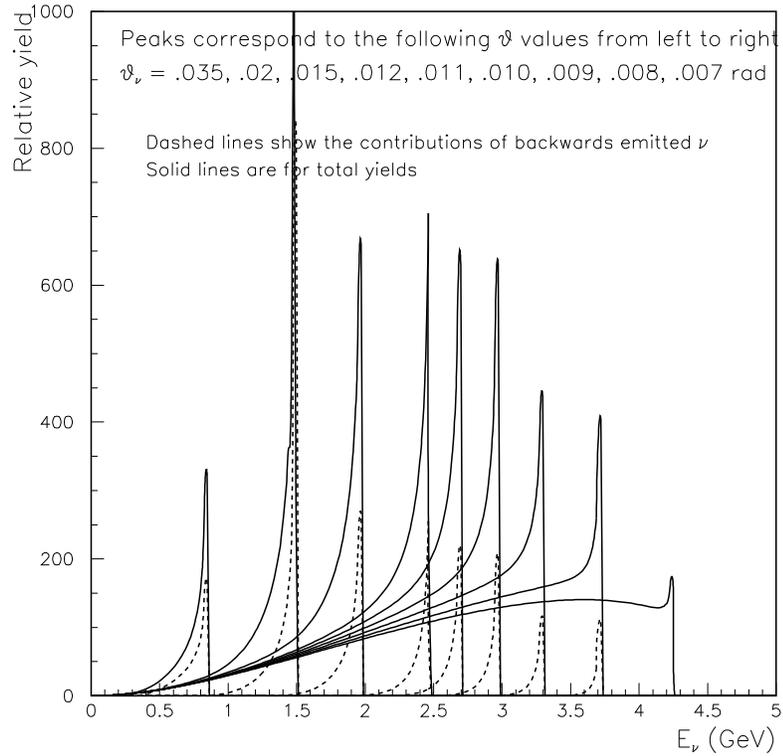,width= .9\linewidth}
\caption{Distribution of $E_{\nu}$ at fixed angle} \label{fig: endist1}
\end{center}
\end{figure}

Equation (\ref{fin}) is the final result of this paper. It shows that there exist in general 
(for $E>E^*$) two contributions to the spectrum of neutrinos emitted at a given laboratory angle, 
provided the meson beam is energetic enough to bring the neutrino emitted backwards in the meson frame to the 
required forward angle. In fig.~\ref{fig: endist1} this second contribution is displayed explicitely.\\



\begin{thebibliography}{99}
\bibitem{McD} Kirk T. McDonald, arXiv:hep-ex/0111033 
\end{thebibliography}
\end{document}